\documentclass[oneside, twocolumn, amsmath, amssymb, nofootinbib, superscriptaddress]{revtex4-1}

\usepackage{cases}
\usepackage{amsmath}
\usepackage{amssymb}
\usepackage{amsfonts}
\usepackage{amssymb}
\usepackage{dcolumn}
\usepackage{bm}
\usepackage{bbm}
\usepackage{graphicx}
\usepackage{xcolor}
\usepackage{array}
\usepackage{subfigure}
\usepackage{hyperref}
\usepackage{wasysym}

\def\l{\left}
\def\r{\right}
\def\be{\begin{equation}}
\def\ee{\end{equation}}
\def\ba{\begin{eqnarray}}
\def\ea{\end{eqnarray}}
\def\bl#1\el{\begin{align}#1\end{align}}

\begin{document}

\title{Ultrahigh-Energy Gamma Rays and Gravitational Waves from Primordial Exotic Stellar Bubbles}

\author{Yi-Fu Cai}
\email{yifucai@ustc.edu.cn}
\affiliation{Department of Astronomy, School of Physical Sciences, University of Science and Technology of China, Hefei, Anhui 230026, China}
\affiliation{School of Astronomy and Space Science, University of Science and Technology of China, Hefei, Anhui 230026, China}
\affiliation{CAS Key Laboratory for Research in Galaxies and Cosmology, University of Science and Technology of China, Hefei, Anhui 230026, China}

\author{Chao Chen}
\email{iascchao@ust.hk}
\affiliation{Department of Astronomy, School of Physical Sciences, University of Science and Technology of China, Hefei, Anhui 230026, China}
\affiliation{Jockey Club Institute for Advanced Study, The Hong Kong University of Science and Technology, Clear Water Bay, Kowloon, Hong Kong, P.R.China}

\author{Qianhang Ding}
\email{qdingab@connect.ust.hk}
\affiliation{Department of Physics, The Hong Kong University of Science and Technology, Clear Water Bay, Kowloon, Hong Kong, P.R.China}
\affiliation{Jockey Club Institute for Advanced Study, The Hong Kong University of Science and Technology, Clear Water Bay, Kowloon, Hong Kong, P.R.China}

\author{Yi Wang}
\email{phyw@ust.hk}
\affiliation{Department of Physics, The Hong Kong University of Science and Technology, Clear Water Bay, Kowloon, Hong Kong, P.R.China}
\affiliation{Jockey Club Institute for Advanced Study, The Hong Kong University of Science and Technology, Clear Water Bay, Kowloon, Hong Kong, P.R.China}

\begin{abstract}

	We put forward a novel class of exotic celestial objects that can be produced through phase transitions occurred in the primordial Universe. These objects appear as bubbles of stellar sizes and can be dominated by primordial black holes (PBHs). We report that, due to the processes of Hawking radiation and binary evolution of PBHs inside these stellar bubbles, both electromagnetic and gravitational radiations can be emitted that are featured on the gamma-ray spectra and stochastic gravitational waves (GWs). Our results reveal that, depending on the mass distribution, the exotic stellar bubbles consisting of PBHs provide not only a decent fit for the ultrahigh-energy gamma-ray spectrum reported by the recent LHAASO experiment, but also predict GW signals that are expected to be tested by the forthcoming GW surveys.

\end{abstract}

\maketitle

\section{Introduction}
With dramatic developments of observational technologies, a large number of new phenomena have been discovered in various astronomical experiments in the past decade. Our understanding of the very nature of Universe, especially of the exotic astrophysical objects including black holes, supernovae, neutrons, blazars, dark matter (DM) and active galactic nuclei, has been greatly developed. Specifically, the high-energy gamma-ray observations provide the only available probe to identify cosmic-ray sources which could tell us the unique information about the exotic celestial objects, as gamma rays travel straightly from the source without the deflection by galactic magnetic field \cite{Kelner:2006tc, Strong:2007nh,Kumar:2014upa,Ge:2020uft}. However, many of cosmic-ray sources have physical origins that are still under discussion \cite{Berezhko:1999sc, Kobayakawa:2000nq}.

Also, accumulated gravitational wave (GW) events detected by the LIGO/Virgo Collaboration started a new era in the observational astronomy, that shed light on the formation of exotic astrophysical objects \cite{GBM:2017lvd}, and have been used to test general relativity in the unexplored strong gravity regime \cite{TheLIGOScientific:2016src}. Hence, the development of the multi-messenger observation, which is a joint observation of cosmic rays, neutrinos, photons and GWs, could provide us the unique insights into the properties of astrophysical sources and source populations in our Universe \cite{Meszaros:2019xej}. With the multi-messenger observation, there is a growing interest in searching for exotic celestial bodies, such as quark stars \cite{Ivanenko:1965dg}, boson stars \cite{Kaup:1968zz}, dark stars \cite{Spolyar:2007qv} and antistar \cite{Dupourque:2021poh}, etc. Their search can reveal new aspects of fundamental science and serve as new probes into the primordial Universe.

In this paper, we propose a novel class of exotic celestial objects filled by primordial black holes (PBHs). These objects appear as bubbles of stellar sizes and can emit gamma rays and GWs through the processes of Hawking radiation and binary mergers, respectively. Theoretically, these stellar bubbles can be generated from some new-physics phenomena that might have occurred in the primordial Universe, such as, quantum tunnelings during or after inflation \cite{Coleman:1980aw, Zhou:2020stj}, multi-stream inflation \cite{Li:2009sp}, inhomogeneous baryogenesis \cite{Cohen:1997ac}, etc. In these cases, before the bubble-wall tension vanishes, the field values are different between inside and outside of the bubble. Such difference can result in different local physics inside the bubble (for PBH cases, see \cite{Belotsky:2018wph, Ding:2019tjk} for details), namely the production rate of the exotic species of matter illustrated below, which indicates the existence of the ``\textit{island universes}'', as a baby version of a \textit{multiverse}.
If the sizes of these bubbles are small enough, say, smaller than the resolution of current telescopes, they behave as exotic celestial objects.

Observational consequence of exotic stellar bubbles can arise from:
(i) Decay, annihilation or interaction of exotic matter inside the bubble, for example, unstable particles, textures, monopoles, cosmic strings, domain walls, and PBHs, etc, that can yield cosmic rays at high energy scales and hence may address the puzzle of ultrahigh-energy cosmic rays, such as PeV gamma rays discovered in the recent LHAASO observation \cite{Cao2021};
(ii) Interaction at the interface of the bubble, for example, if the stellar bubble is made of antimatter, where the effects can be in analogy to \cite{Dupourque:2021poh}. If there are overdense regions inside the bubble, certain amount of PBHs can be formed due to the local gravitational collapse \cite{Zeldovich:1966, Hawking:1971ei, Carr:1974nx}, resulting in a wide range of PBH masses which is quite different from the astrophysical black holes \cite{Khlopov:2008qy, Sasaki:2018dmp, Carr:2016drx}. When these PBHs' masses are small enough, their Hawking radiations become significant to generate observable electromagnetic (EM) signals. Additionally, PBHs can also cluster to form binaries and generate GWs from their mergers \cite{Bird:2016dcv, Sasaki:2016jop}.
As such, it is expected to search for the nontrivial observational signals from these exotic celestial bodies, as we will investigate in this paper.

\section{EM radiations}
In \cite{Hawking:1974rv, Hawking:1974sw} Hawking found remarkably that black holes emit radiations in the black-body form. Theoretically, the PBHs satisfy a certain mass distribution depending on the underlying generation mechanisms \cite{Cai:2018tuh, Chen:2019zza, Carr:2018poi, Zhou:2020kkf}, and hence one can derive the EM radiations contributed by all possible mass scales. Let us focus our interest on neutral PBHs of the Schwarzschild type. The time-dependent physical number density $n_i$ of elementary particle $i$ emitted by a distribution of PBHs per unit time and per unit energy can be determined by
\be \label{Hawk_tot}
{\mathrm{d}^2 n_i \over \mathrm{d}t \mathrm{d}E}(E) = \int_{M_\text{min}}^{M_\text{max}} \frac{\mathrm{d}^2 N_i}{\mathrm{d}t \mathrm{d}E}(E,M) n_\text{PBH}(M) \mathrm{d}M ~,
\ee
where $\mathrm{d}^2 N_i / \mathrm{d}t \mathrm{d}E$ is the Hawking instantaneous emission rate for particle $i$. The lower mass limit $M_\text{min}$ is usually set to the Planck mass $M_\text{pl} \simeq 2 \times 10^{-5}$ g, while the upper one $M_\text{max}$ can be infinity.

In general, the mass distribution of PBHs can be described by the differential physical mass function $n_\text{PBH}(M) \equiv \mathrm{d} n_\text{PBH} / \mathrm{d} M$, where $\mathrm{d} n_\text{PBH}$ is the physical number density of PBHs in the mass range $(M, M + \mathrm{d}M)$. For heavy PBHs which survive at present (the formation mass $M_f \gtrsim 10^{15}$g), the current energy fraction $\tilde{f}_{\text{PBH}}$ of PBHs over the DM component inside PBH bubbles is given by the integral $\tilde{f}_{\text{PBH}} \equiv \tilde{\Omega}_\text{PBH} / \tilde{\Omega}_\text{DM} = \int_{M_\text{min}}^{M_\text{max}} \psi(M) \mathrm{d}M$, where $\tilde{\Omega}_\text{PBH}$ and $\tilde{\Omega}_\text{DM}$ are the current normalized density of PBHs and DM inside PBH bubbles, respectively. Note that, we use the notations $\Omega_{\text{rad}}$ and $\Omega_{\text{DM}}$ to denote the corresponding quantities defined over the whole observable Universe. The current mass function $\psi(M)$ is defined as $\psi(M) \equiv M n_\text{PBH}(M) / \rho_\text{DM}$, where the scale factor at present takes $a(t_0) = 1$. If the PBHs were from inflationary fluctuations, they naturally satisfy a lognormal mass function \cite{Dolgov:1992pu, Carr:2016drx},
\be \label{LN_psi}
\psi_\text{LN}(M) = {\tilde{f}_{\text{PBH}} \over \sqrt{2 \pi} \sigma M} \exp\Big[ - { \ln^2(M/M_\text{pk}) \over 2 \sigma^2 } \Big] ~,
\ee
where the subscript ``LN" denotes the lognormal type. Given $\tilde{f}_{\text{PBH}}$, there are two parameters in $\psi_\text{LN}(M)$: $\sigma$ describes the width of mass distribution and $M_\text{pk}$ is the mass at which the function $M \psi_\text{LN}(M)$ (the DM faction in PBHs at the logarithmic interval around $M$) peaks. Since light PBHs ($M_f \lesssim 10^{15}$ g) would have evaporated at earlier time, the present mass function could be deformed from the initial shape, especially in the low-mass tail \cite{Carr:2016hva}.

Let us emphasize that due to the clustering of PBHs inside PBH bubbles, the local density $\tilde{f}_{\text{PBH}}$ can be amplified. Taking the multi-stream inflation as an example, we use $\beta_1$ to denote the volume fraction of PBH bubbles in the observed Universe. If we assume the dark matter energy density is same inside PBH bubbles and the whole observable Universe, i.e., $\Omega_{\text{DM}} = \tilde{\Omega}_{\text{DM}}$, one has the following relation: $\tilde{f}_{\text{PBH}} = \beta_1^{-1} f_{\text{PBH}}$, where $f_{\text{PBH}} \equiv \Omega_{\text{rad}}/\Omega_{\text{DM}}$ is the standard definition. In the multi-stream inflation, the probability $\beta_1 \ll 1$, so that $\tilde{f}_{\text{PBH}}$ is largely amplified. According to the current constraints on $f_{\text{PBH}}$ \cite{Carr:2020xqk}, $f_{\text{PBH}} \sim \mathcal{O}(10^{-3})$ for the wide mass window $10^{-16} - 10^{1} M_\odot$, it is straightforward to know that $\tilde{f}_{\text{PBH}} > 1$ requires $\beta_1 < 10^{-3}$, corresponding to the size of PBH bubbles: $V_\text{bubb} < 3$ Mpc, which is quite easily satisfied in the multi-stream inflation. Actually, for a stellar bubble, the amplification could be much larger than $10^3$. The number of PBHs in the clustering can be estimated by
\be
N = \left( \frac{k_2}{k_1} \right)^3 \tilde{\beta} ~,
\ee
Here, $1/k_1$ denotes the comoving scale which exits the horizon at the time of bifurcation in the multi-stream inflation and $1/k_2$ denotes the comoving scale which exits the horizon at the time of PBH formation during inflation. $\tilde{\beta}$ is the initial PBH abundance inside a PBH bubble that corresponds to the current local density $\tilde{f}_{\text{PBH}}$. The scale $1/k_1$ determines the size the exotic stellar bubble, which depends on the potential of the multi-stream inflation. In practice, we take $k_1^{-1} = 1 \mathrm{Mpc}$ as an example. $k_2$ is related to the mass of PBHs as \cite{Nakama:2016gzw}
\begin{align}
k_2 \simeq 7.5 \times 10^5 \mathrm{Mpc}^{-1}\left( \frac{M_\mathrm{PBH}}{30 M_{\odot}} \right)^{-1/2} ~,
\end{align}
where $M_{\odot}$ is the solar mass. We take $M_\mathrm{PBH} = 10^{15} \mathrm{g}$, which gives $k_2 \sim \mathcal{O}(10^{15}) \mathrm{Mpc}^{-1}$ and the number of PBHs in the clustering is estimated as
\begin{align}
N \simeq 2 \times 10^{47} \tilde{\beta} ~,
\end{align}
The calculation of PBH abundance in PBH stellar bubbles is similar to the standard calculation \cite{Carr:2016drx}. For example, for the critical collapse model \cite{Niemeyer:1997mt, Carr:2020xqk, Luo:2020dlg}, the local abundance is given by $\tilde{\beta} \simeq k \sigma^{2 \gamma} \text{erfc} \left(\frac{\delta_c}{\sqrt{2} \sigma}\right)$. Hence, $N > 10^{17}$ if $\tilde{\beta} > 10^{-30}$, which is easily realized in various PBH formation mechanisms \cite{Carr:2020xqk}. Thus, the number $N$ of PBHs inside a PBH bubble in general are largely amplified, it is safe to regarded $N$ as a free parameter in the following discussions.

The extragalactic gamma-ray background (EGB) is an important constraint for evaporating PBHs \cite{Page:1976wx}. With the advent of gamma-ray experiments in various energy ranges, such as the Imaging Compton Telescope (COMPTEL) for $0.8 \sim 30~$MeV \cite{weidenspointner2000cosmic}, the Energetic Gamma Ray Experiment Telescope (EGRET) for $20~$MeV $\sim 30~$GeV \cite{Strong:2004ry}, the Fermi Large Area Telescope (Fermi LAT) for $20~$MeV $\sim 300~$GeV \cite{Atwood:2009ez}, PBHs within the mass range $10^{14} \sim 10^{16}$ g are severely constrained and cannot provide the dominant contribution to DM \cite{MacGibbon:1991vc, Carr:1998fw, Carr:2009jm}. Moreover, the galactic gamma-ray background \cite{Carr:2016hva} and gamma-ray bursts \cite{Ukwatta:2015iba, Jung:2019fcs} can be applied to constrain PBHs.

\begin{figure}[ht]
	\centering
	\includegraphics[width=\linewidth]{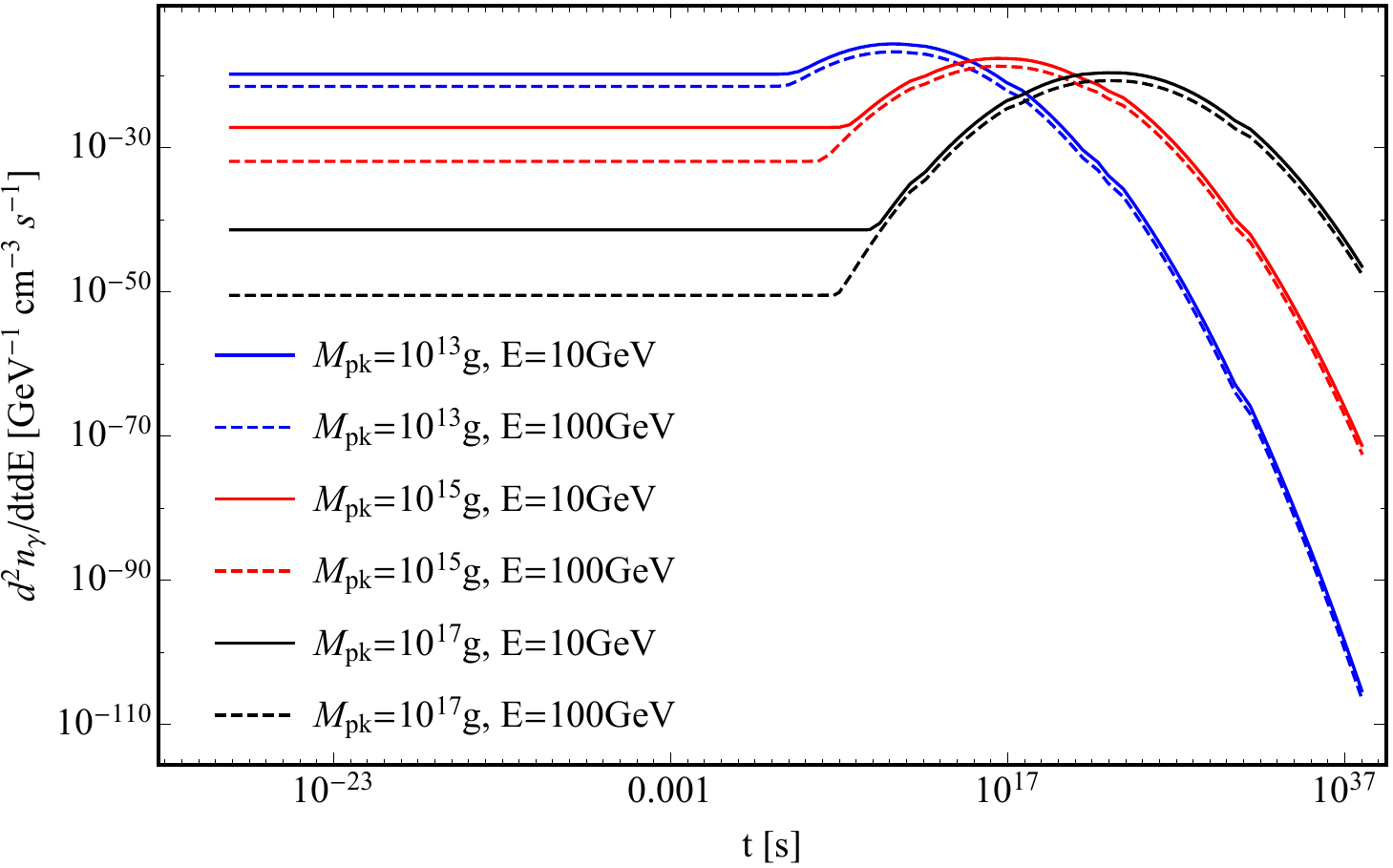}
	\caption{The spectrum $\mathrm{d}^2 n_\gamma / \mathrm{d}t \mathrm{d}E$ of a PBH bubble as a function of cosmic time $t$ for peak masses $M_\text{pk} = 10^{13}, 10^{15}, 10^{17}$g and photon energies $E = 10, 100~$GeV, respectively. For simplicity, we have set $\tilde{f}_{\text{PBH}} =1$ and $\sigma=1$.
	}
	\label{fig:radiation1}
\end{figure}

\begin{figure}[ht]
\centering
\includegraphics[width=\linewidth]{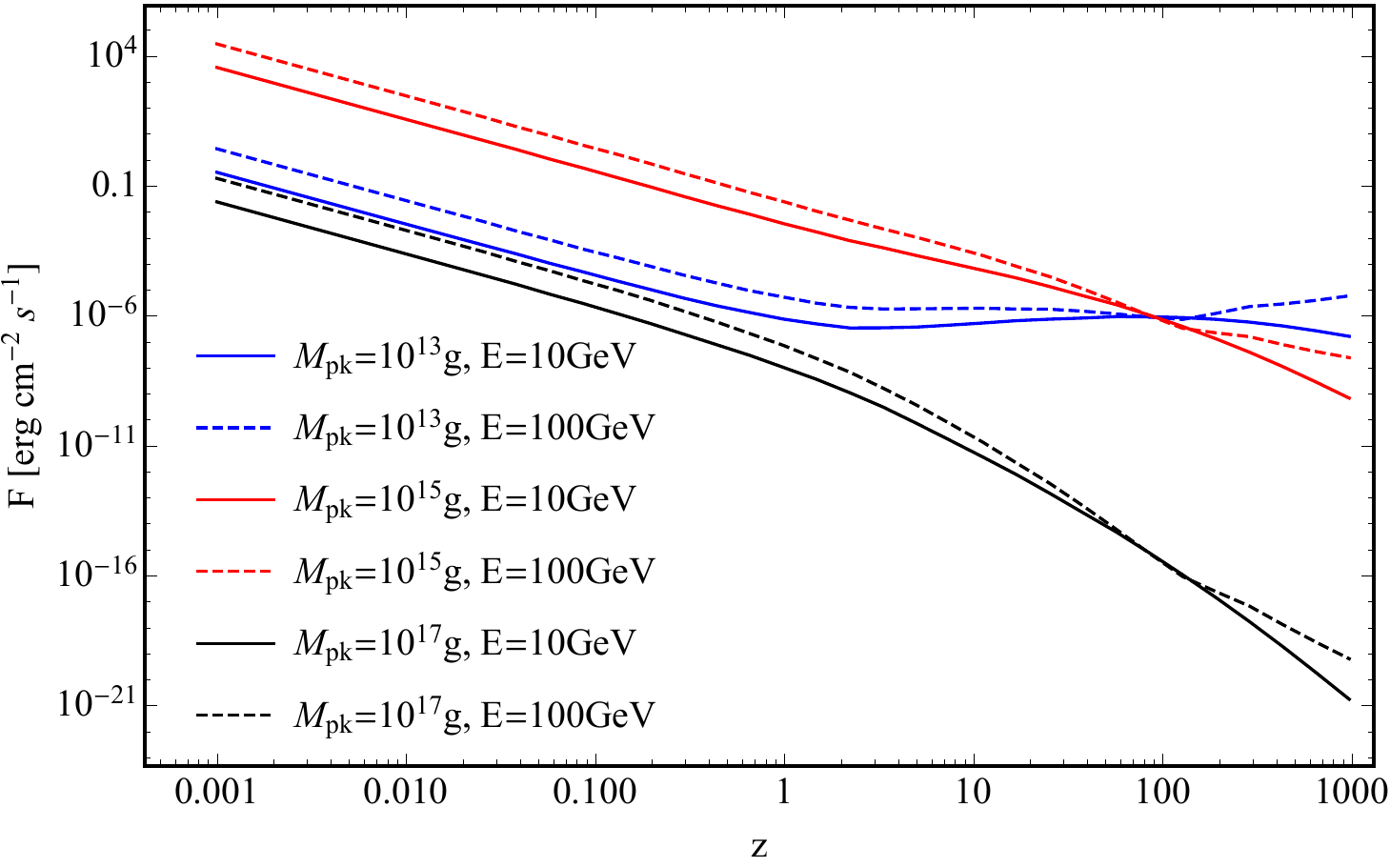}
\caption{
	The observed photon flux $F$ of a PBH bubble of physical volume $1$ Mpc$^3$ located at different redshifts, for various peak masses and photon energies.
}
\label{fig:radiation2}
\end{figure}

Now, we study the EM signals from a single PBH stellar bubble and the detectability in gamma-ray channels. Consider an initial lognormal distributed PBH stellar bubble located at redshift $z$. The key EM observable is the photon flux detected on Earth $F(\tilde{E},t) \equiv \tilde{E}^2 \mathrm{d}^2\tilde{n}_\gamma / \mathrm{d} \tilde{E} \mathrm{d}t$, where $\tilde{E}$ and $\tilde{n}_\gamma$ are the photon energy and number density observed on Earth. This observed photon flux is related to the intrinsic luminosity $L(E,z)$ of Hawking radiation from the PBH bubble located at $z$ via
\be \label{flux}
F(E,z) = { L(E(1 + z),z) \over 4 \pi d_L^2(z) } ~,
\ee
where $L(E,z)$ is determined by the emitted photon physical number density per unit energy and per unit time $\mathrm{d}^2 n_\gamma / \mathrm{d}t \mathrm{d}E$.
It can be numerically computed as shown in Fig.~\ref{fig:radiation1} for the peak masses $M_\text{pk} = 10^{13}, 10^{15}, 10^{17}$g and the energies of emitted photons $E = 10, 100~$GeV, respectively. The flat parts in this panel are caused by small masses radiating at the early stage of Hawking radiation.
The luminosity distance in Eq.~\eqref{flux} is given by $d_L(z) = (1 + z) \int_0^z \mathrm{d}\tilde{z} / H(\tilde{z})$, which accounts for the redshift of the photon energy and apparent emission rate \cite{Dodelson:2003ft, Mukhanov:2005sc}. Combing \eqref{Hawk_tot} and \eqref{flux} yields the observed photon fluxes in various photon energy ranges from a single PBH bubble located at $z$, as reported in Fig.~\ref{fig:radiation2}. It shows that for the small peak mass $M_\text{pk} = 10^{13}$g, the observable fluxes of the PBH bubble of redshift $z\gtrsim1$ would increase to some extent in contrast to the larger peak mass $M_\text{pk} = 10^{15}, 10^{17}$g. Due to the evaporation of a majority of PBH bubbles at high redshifts.

\begin{figure}[ht]
	\centering
	\includegraphics[width=\linewidth]{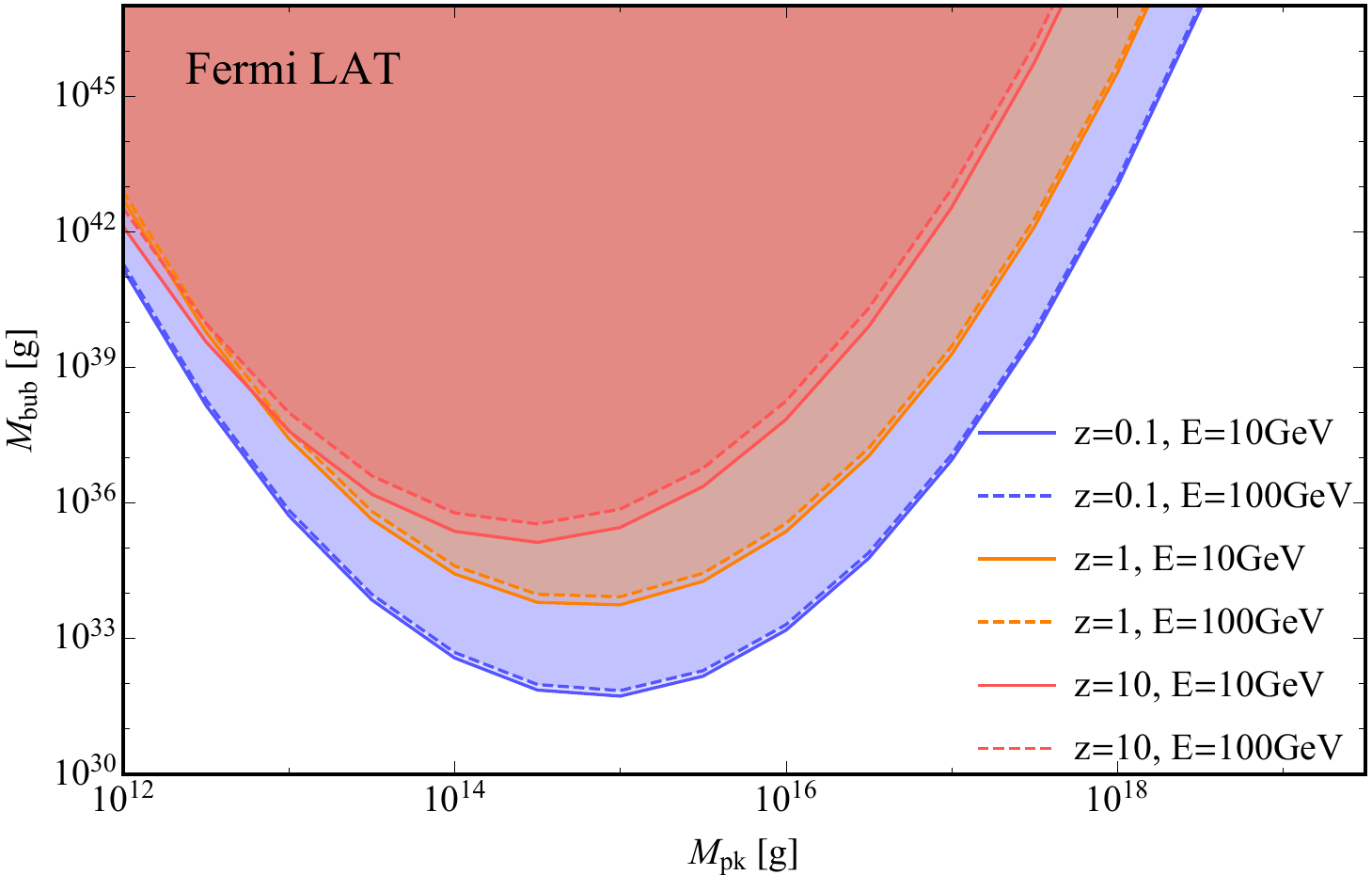}
	\caption{
		The parameter space of the peak mass $M_\text{pk}$ and bubble mass $M_\text{bub}$ of a PBH bubble allowed by Fermi LAT (the shaded regions) for redshifts $z=0.1, 1, 10$ and photon energies $E= 10, 100~$GeV, respectively.
	}
	\label{fig:radiation3}
\end{figure}

\begin{figure}[ht]
	\centering
	\includegraphics[width=\linewidth]{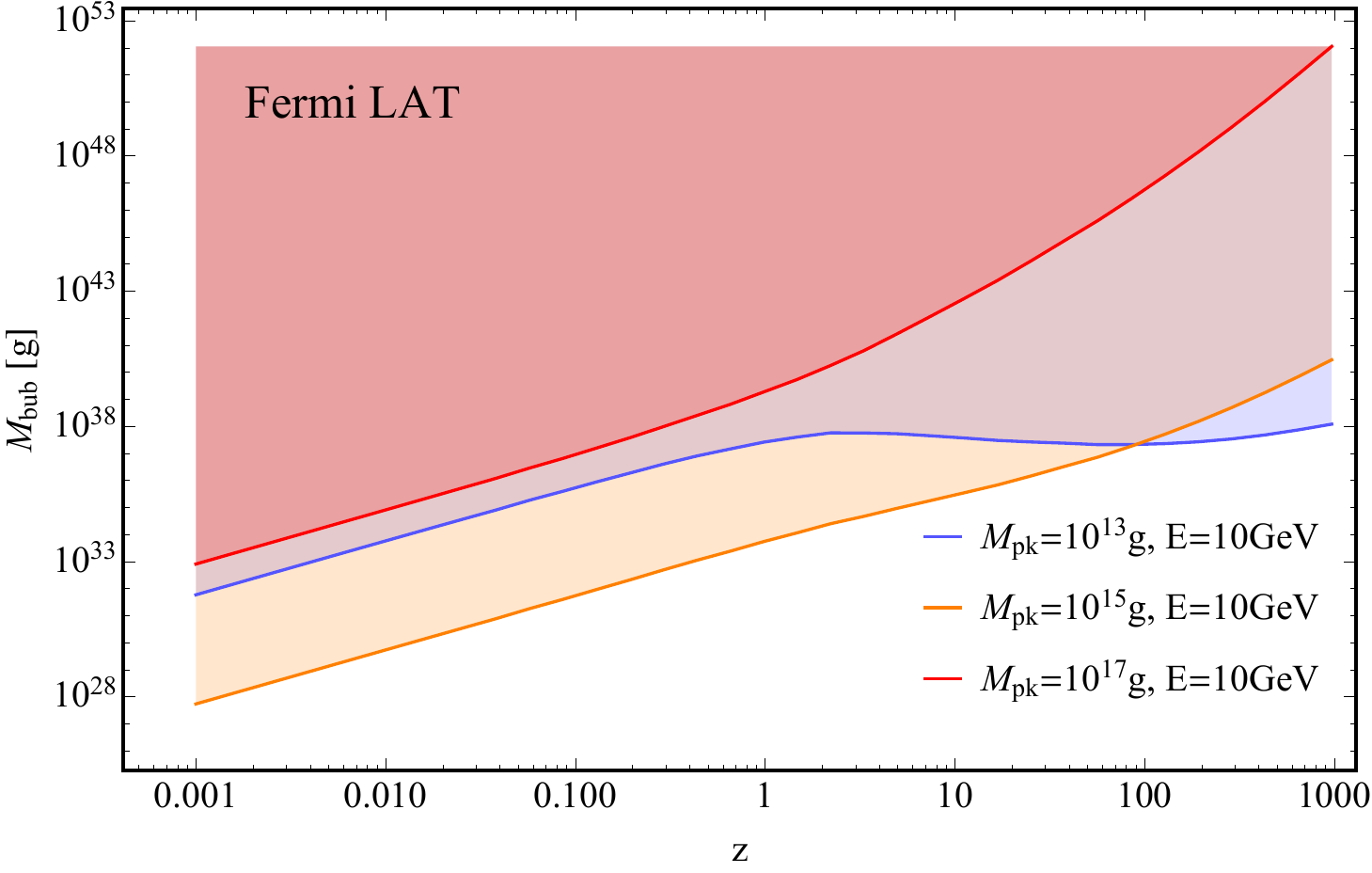}
	\caption{ 
		The parameter space of the redshift $z$ and bubble mass $M_\text{bub}$ allowed by Fermi LAT (the shaded regions) with the photon energy $E = 10~$GeV and peak masses $M_\text{pk} = 10^{13}, 10^{15}, 10^{17}$g, respectively.
	}
	\label{fig:radiation4}
\end{figure}

\section{Gamma-ray signals}
Recall that the PBH mass function in \eqref{LN_psi} relies on three parameters, i.e., $\tilde{f}_{\text{PBH}}$, $\sigma$ and $M_\text{pk}$. In the case that PBHs cluster as a single bubble, we generalize the concept of $\tilde{f}_{\text{PBH}}$ to be the energy density of PBHs inside the bubble, which is convenient for our calculation. Furthermore, the intrinsic luminosity of Hawking radiation involves two other parameters of a bubble, i.e., its physical volume $V$ and redshift $z$. For observable interest, we combine $\tilde{f}_{\text{PBH}}$ and $V$ to yield the initial total mass of a PBH bubble $M_\text{bub}$. Thus, we get four parameters: $M_\text{bub}$, $z$, $M_\text{pk}$ and $\sigma$.
Using the point-source differential sensitivity in the 10-year observation of Fermi LAT for a high Galactic latitude (around the north Celestial pole) source \cite{Atwood:2009ez}, we numerically derive the experimentally allowed parameter spaces for ($M_\text{bub}$, $M_\text{pk}$) and ($M_\text{bub}$, $z$) by the shaded regions in Fig.~\ref{fig:radiation3} and Fig.~\ref{fig:radiation4}, respectively. Setting $\sigma=1$, for given redshifts and photon energies, the lower bound of the parameter space for $M_\text{pk}$ and $M_\text{bub}$ is around $M_\text{pk} \simeq 10^{15}$g, which is the mass scale evaporating at present, and the corresponding bubble mass is $M_\text{bub} \simeq 10^{32} \text{g}$. For the smaller or larger values of $M_\text{pk}$, the corresponding Hawking radiation has either decayed out already or not yet become efficient, and thus the bubble mass $M_\text{bub}$ needs to be heavy enough to yield observable evidences. Additionally, Fig.~\ref{fig:radiation3} and Fig.~\ref{fig:radiation4} indicate that, the closer the stellar PBH bubbles are to the Earth, the easier they could be probed. The ``plateau" for the case of $M_\text{pk} = 10^{13}$g and $z \gtrsim 1$ appeared in Fig.~\ref{fig:radiation4} is due to the same reason for the fluxes of $M_\text{pk} = 10^{13}$g shown in the upper right panel.

\begin{figure}[ht]
	\centering
	\includegraphics[width=\linewidth]{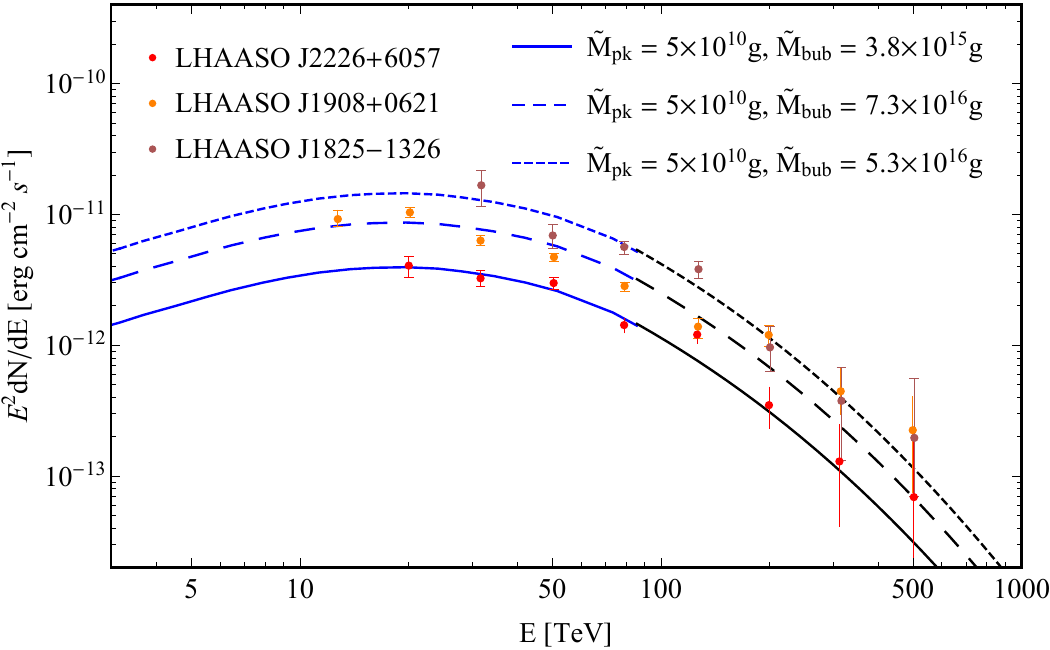}
	\caption{ The fit of three different PBH stellar bubbles to the data by LHAASO. The observation distance of ultrahigh-energy gamma-ray sources $\mathrm{J}2226+6057$, $\mathrm{J}1908+0621$, $\mathrm{J}1825-1326$ are $0.8$, $2.37$, $1.55$ $\mathrm{kpc}$ \cite{Kothes:2001px, Chen:2021hen}, respectively. The blue curves are numerical results produced using the publicly available code BlackHawk \cite{Arbey:2019mbc}, and the black curves are the superpositions of analytic black body spectra normalized by the BlackHawk calculation.
	}
	\label{fig:LHAASO1}
\end{figure}

Recently, the LHAASO experiment reported the astonishing detection of twelve ultrahigh-energy gamma-ray sources \cite{Cao2021}, which indicates some high energy physics in stellar objects. While the astrophysical sources responsible for these events are under debate \cite{Kelner:2006tc, Strong:2007nh, Albert:2021vrd}, it deserves to examine the possibility of their origins being exotic celestial objects including the PBH stellar bubble. Therefore, we confront our scenario with the latest LHAASO data and report the numerical results in Fig.~\ref{fig:LHAASO1}.
For connecting the gamma-ray spectrum and the present PBH mass distribution, the present lognormal distribution is considered instead of a primordial one. Accordingly, $\tilde{M}_\text{pk}$ denotes the peak mass of present lognormal distribution, and $\tilde{M}_{\mathrm{bub}}$ denotes the present stellar bubble mass. In the numerical calculations, we set $\sigma = 1$. Varying $\sigma$ may lead to better fits to observations, which we leave to future work.
Our results illustrate well that the PBH stellar bubbles can provide a decent fit as shown in the plot explicitly.

\begin{figure}[ht]
	\centering
	\includegraphics[width=\linewidth]{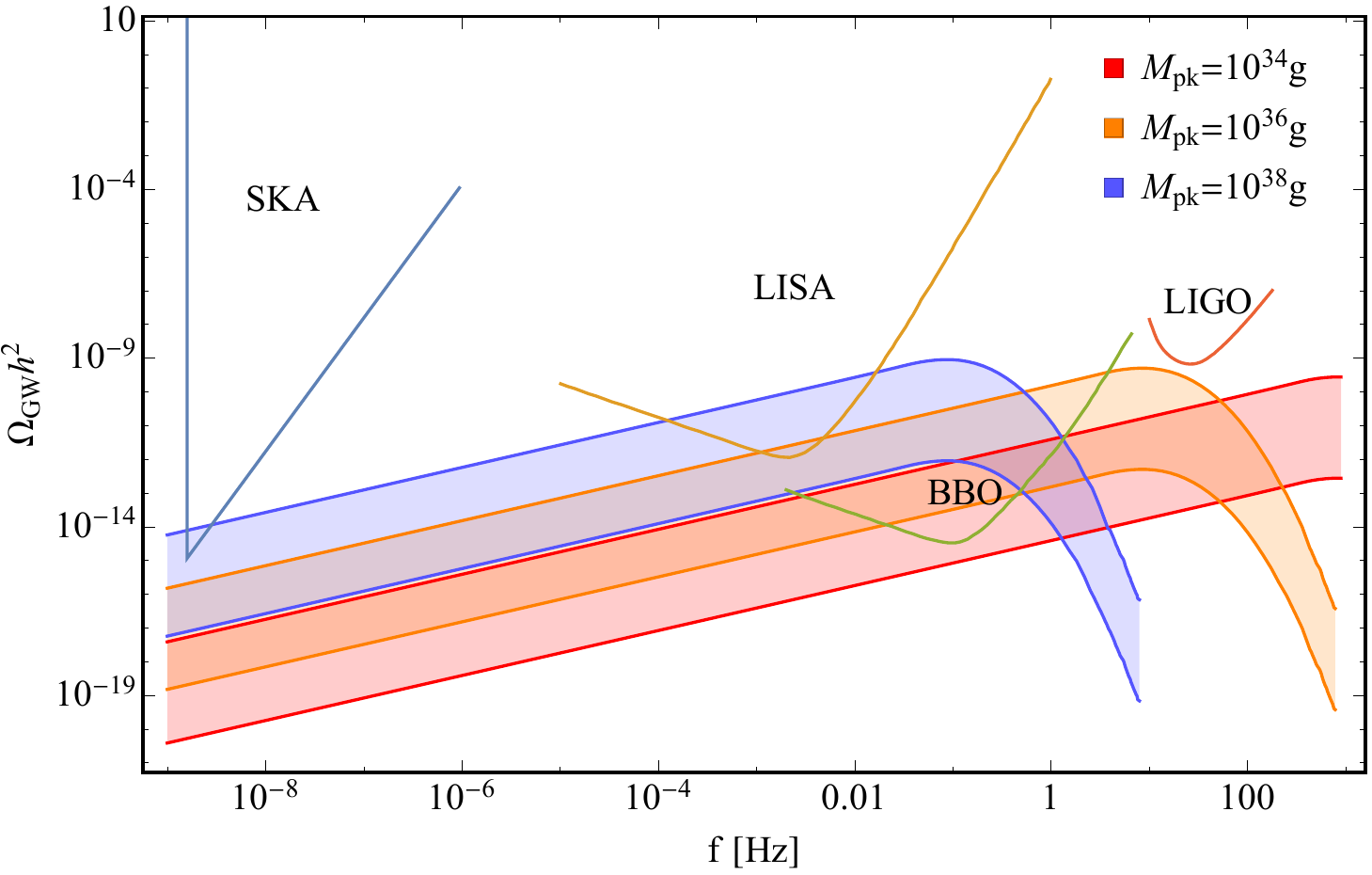}
	\caption{ The GW spectrum of a PBH stellar bubble. We set $\tilde{f}_{\text{PBH}} = 1$ in the bubble and $z = 0.01$. The mass distribution of PBHs is lognormal with $\sigma = 1$ and $M_\text{pk} = 10^{34}, 10^{36}, 10^{38}\mathrm{g}$ for red, orange, blue shadow regions respectively. The lower and upper solid curves denote the total PBH mass in a bubble, which are set as $10^{45} \mathrm{g}$ and $10^{48} \mathrm{g}$, respectively. Sensitivity curves of SKA \cite{2009IEEEP..97.1482D}, LISA \cite{bender1998lisa}, BBO \cite{Crowder:2005nr} and LIGO \cite{Abbott:2007kv} are plotted.
	}
	\label{fig:LHAASO2}
\end{figure}

\begin{figure}[ht]
	\centering
	\includegraphics[width=\linewidth]{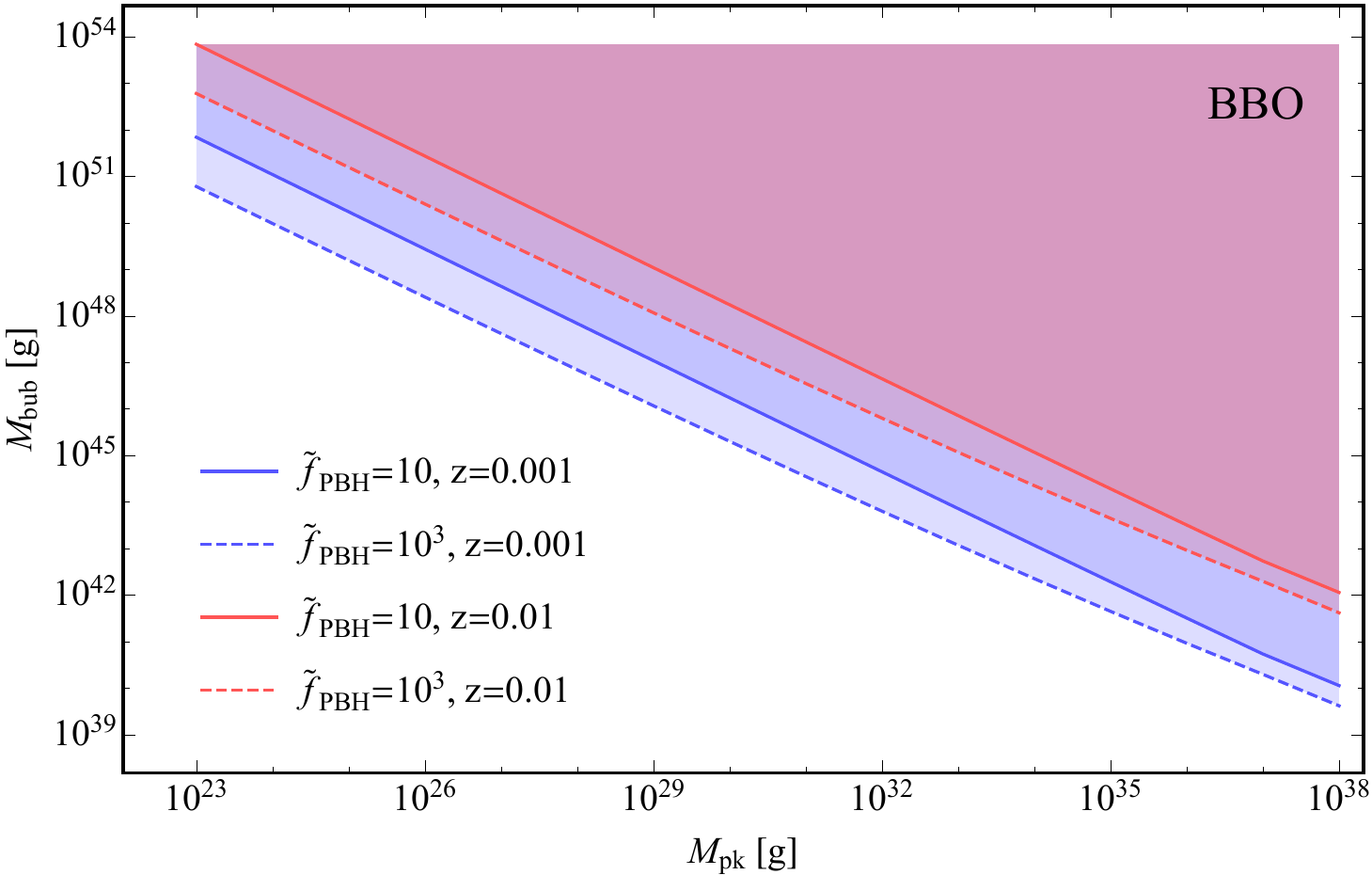}
	\caption{
		The parameter space of the bubble mass and peak mass of a PBH bubble that can be probed by BBO at $0.1 \mathrm{Hz}$ with $\Omega_{\mathrm{GW}}h^2 = 2.9 \times 10^{-15}$. The solid and dashed lines denote $\tilde{f}_{\text{PBH}} = 10, 10^3$, respectively. Blue and red shaded regions denote $z = 0.001$, $0.01$, respectively.
	}
	\label{fig:LHAASO3}
\end{figure}

\section{GW signals}
The evolution of PBH binaries produces GWs. For those PBHs whose masses are less than $10^4 M_{\odot}$ and the associated GWs become very weak for observations. Accordingly, the main observational channel of GWs is to follow the evolution of PBH binaries when considering heavy masses. A PBH binary in a stellar bubble forms from two nearby black holes and decouples from the Hubble flow. Given the initial separation $x$ in a binary with total mass $M = m_1 + m_2$, the binary system forms at $z \approx 3(1 + z_{\mathrm{eq}})/\lambda - 1$, where $z_{\mathrm{eq}}\approx 3000$ is the redshift at matter-radiation equality in the $\Lambda \mathrm{CDM}$ cosmology with $\Omega_{\mathrm{M}} = 0.315$, $\Omega_{\Lambda} = 0.685$ \cite{Aghanim:2018eyx}, and $\lambda \equiv 8 \pi \rho_{\mathrm{eq}} x^3/3 M$ \cite{Ali-Haimoud:2017rtz}.

The GW energy spectrum from a single PBH stellar bubble is
\be \label{GWenergy}
	\Omega_{\mathrm{GW}}(f) = \frac{1}{\rho_c}\frac{1}{4 \pi d_L^2} f_r \frac{dE_{\mathrm{GW}}}{df_r} R~,
\ee
where $f$ and $f_r$ are the GW frequency in observed frame and binaries rest frame, with $f = f_r/(1 + z)$. $\rho_{c} = 3 H_0^2/8 \pi G$ the critical density of the Universe, and $R$ is the comoving merger rate of PBH binaries \cite{Chen:2018czv}. $dE_{\mathrm{GW}}/df_r$ is the energy emission per frequency interval, which is parameterized by \cite{Cutler:1993vq, Chernoff:1993th, Zhu:2011bd}
\be
\begin{aligned}
 \label{dEnergy}
	&\frac{dE_{\mathrm{GW}}}{df_r} 
	\\=& (\pi G)^{2/3}\mathcal{M}_c^{5/3}
	\\& \times
	\left\{
	\begin{aligned}
		&f_r^{-1/3} \qquad  ,~f_r < f_1 \\
		&f_r^{2/3} f_1^{-1}  \qquad ,~f_1 \leq f_r <f_2 \\
		&f_4^4 f_r^2\big[ f_1 f_2^{4/3} \big( 4(f_r - f_2)^2 + f_4^2 \big)^2 \big]^{-1},~f_2 \leq f_r < f_3
	\end{aligned}
	\right.
\end{aligned}
\ee
where $\mathcal{M}_c \equiv (m_1 m_2)^{3/5}/M^{1/5}$ is the chirp mass of the binary system and $f_i = (a_i \eta^2 + b_i \eta +c_i)/\pi G M$. The symmetric mass ratio $\eta$ is defined as $\eta \equiv m_1 m_2/M^2$ and the coefficients $a_i,~b_i,~c_i$ can be found in Table $1$ of \cite{Ajith:2007kx}. In Eq.~\eqref{dEnergy} the emitted GW energy is proportional to $\mathcal{M}_c^{5/3}$, and thus the contributions of light PBH binaries to GWs are negligible.

Combining \eqref{GWenergy} and \eqref{dEnergy} yields the GW spectrum from a single PBH stellar bubble and detectable mass parameters region in Fig.~\ref{fig:LHAASO2} and Fig.~\ref{fig:LHAASO3}, respectively.
In Fig.~\ref{fig:LHAASO2}, with the same $M_\mathrm{bub}$, the larger $M_\mathrm{pk}$, more GWs are emitted at the same frequency. Bubbles with peak mass around $10^{15} \mathrm{g}$ emit weak GWs that are difficult to be probed within $(10^{-9}, 10^{2}) \mathrm{Hz}$, which is also shown in Fig.~\ref{fig:LHAASO3}, where the $M_\mathrm{bub}$ should be larger than $10^{53} \mathrm{g}$ with $M_\mathrm{pk}$ less than $10^{23} \mathrm{g}$ at $0.1 \mathrm{Hz}$ for $\tilde{f}_{\mathrm{PBH}} = 10^3$ and $z = 0.01$.

While PBH stellar bubbles with strong Hawking radiation can hardly be probed by the current GW surveys, they may be detectable in some ultra-high frequency GW experiments. Note that the peak energy density in GW spectrum with different peak mass is similar. In Eq.~\eqref{dEnergy}, the peak energy radiation occurs at $f_r = f_2$. Comoving merger rate of binaries scales as $R \sim M^{-32/37}$, see \cite{Sasaki:2016jop, Chen:2018czv, Ding:2019tjk} for details. Thus, the total radiation power in binaries' rest frame is $R f_r dE_{\mathrm{GW}}/df_r \sim M^{5/37} \eta (a_2 \eta^2 + b_2 \eta +c_2)^{5/3}/(a_1 \eta^2 + b_1 \eta +c_1)$, which depends on symmetric mass ratio $\eta$ and weakly depends on $M$. Due to the $M_\text{pk}$ independence in $\eta$ distribution, the peak energy density in GW spectrum is weakly dependent on the peak mass in PBH mass distribution. This provides the possibility of detecting GW signals of strong Hawking radiation PBH stellar bubble in ultra-high frequency range.

Apart from the GW signals from the PBH binaries inside PBH stellar bubbles as we have discussed above, the dynamics of bubbles can also produce the stochastic GW background, e.g., the bubble nucleation through the first-order phase transition, bubble expansion and the collision of bubbles \cite{Cai:2017cbj}. Assuming PBH bubbles as a set of isolated spherical bubbles in this work, which can not produce GWs through nucleation and expansion process. The dominated stochastic GW background from the bubble dynamics channel is therefore from the collision of bubbles. As for the phase transition during the inflation, the scalar field plays an essential role in producing GW background during the collision of bubbles \cite{PhysRevD.45.4514, PhysRevLett.69.2026}. The resulting GW energy spectrum $\Omega_{\mathrm{GW}}$ follows $\Omega_{\mathrm{GW}}(f) \propto f^q$, where the spectral indices $q = 2.8, -1$ in the limits of low and high frequencies, respectively \cite{Huber:2008hg}. While for the phase transition occurs after the inflation, the GW signals are also contributed by the sound waves \cite{Hindmarsh:2013xza} and magnetohydrodynamic (MHD) turbulence \cite{Caprini:2006jb} in the plasma. The GW energy spectral indices from sound waves in the plasma take $q = 3, -3$ for the low and high frequencies, respectively \cite{Hindmarsh:2015qta}. While the MHD turbulence produces the GWs with spectral indices $q = 3, -5/3$ in the low and high frequencies, respectively \cite{Caprini:2009yp}. Hence, the GW signals produced from the bubble dynamics can be distinguished from the signals from the PBH binary mergers inside the PBH bubbles by using the future GW observations (see Refs. \cite{Caprini:2015zlo, Binetruy:2012ze} and references therein). The potential effects of the dynamics of PBH bubbles on GW signals are needed to investigate further in the follow-up work.

\section{Concluding remarks}
To conclude, we propose the hypothetical possibility of stellar bubbles, which are star-like objects in the sky with exotic features. We focus on a specific class of exotic stellar bubbles filled by lognormal distributed PBHs and analyze their signatures through both the EM and GW observational windows. For the EM channel, a bubble dominated by light PBHs can yield detectable gamma-ray spectra via Hawking radiation. The peak mass $M_\text{pk} \sim 10^{15}$g and bubble mass $M_{\mathrm{bub}} \sim 10^{32}$g can be related to the $10-100~$GeV detection band of Fermi LAT (Fig.~\ref{fig:radiation3} and Fig.~\ref{fig:radiation4}). Impressively, this scenario can make a decent fit to the ultrahigh-energy gamma-ray events discovered by LHAASO and hence hint to the existence of PBH stellar bubbles with the present peak mass $\tilde{M}_\text{pk} \sim 10^{10} \mathrm{g}$ (see Fig.~\ref{fig:LHAASO1}). For the GWs channel, we find that massive PBH binaries with $M_\text{pk} \sim 10^{34} - 10^{38}$ g and $M_{\mathrm{bub}} \sim 10^{45} - 10^{48}$g can produce detectable GWs within in the frequency band of LISA and BBO (see Fig.~\ref{fig:LHAASO2}).

We comment that, EM and GW signals are complementary for light and heavy bubbles.
	The search for light PBH bubbles can be promising by observing amount of gamma-ray sources whose spectra follow Hawking spectra.
For heavy bubbles, they can be more accessible by the GW astronomy. These combined limits can further constrains the mass function of the PBHs, and thus infer the PBHs formation in the Universe. Accordingly, if such PBH stellar bubbles were observed, it serves as a novel window to probe the very early Universe.

Additionally, cosmic neutrinos and ultra-high frequency GWs could also be generated from these exotic stellar bubbles. The IceCube Neutrino Observatory has confirmed the high-energy cosmic neutrinos as the key messengers where wavelengths are opaque to EM signals \cite{Aartsen:2013jdh}. On the other hand, the PBH bubbles may leave significant GWs of extremely high frequencies, which urges the development of ultra-high frequency GW technology \cite{Aggarwal:2020olq}.
We end by mentioning that, instead of PBH stellar bubbles, there exist other types of bubbles that produce observable signals, such as collision of cosmic strings \cite{Shellard:1987bv} and domain walls \cite{Takamizu:2004rq} in the bubbles, matter-antimatter annihilation on boundaries \cite{Blinnikov:2014nea}, etc. These exotic stellar bubbles from the primordial Universe can unveil rich physics hidden in the light of stars, which deserve to be explored in the future.

\section*{Acknowledgement} 
We are grateful to Xiaojun Bi, Ruoyu Liu and Ruizhi Yang for valuable discussions. 
YFC and CC are supported in part by the NSFC (Nos. 11653002, 11961131007, 11722327, 11421303), by the National Youth Talents Program of China, by the Fundamental Research Funds for Central Universities, by the CSC Innovation Talent Funds, and by the USTC Fellowship for International Cooperation.
QD and YW are supported in part by the CRF grant C6017-20GF by the RGC of Hong Kong SAR, and the NSFC EYS (Hong Kong and Macau) Grant No. 12022516.
We acknowledge the use of the clusters {\it LINDA} and {\it JUDY} of the particle cosmology group at USTC and computing facilities at HKUST.

\appendix

\section{Abundance and size of the stellar bubbles} 
The abundance and size of the bubbles depend on the early universe mechanisms. In general, the abundance is determined by the probability of the tunneling (for phase transition) or bifurcation (for multi-stream inflation). And the size of the bubble is determined by the comoving scale at which tunneling or bifurcation happened. For example:

Multi-stream inflation: The radius of the bubble is similarly $R_b = R_0\exp(-N_b)$, where $N_b$ is interpreted as the e-folding number between the beginning of the observable inflation to the bifurcation. Since the bifurcated path eventually merge, the tension of the bubble wall vanishes automatically. The number density of the bubble $n_b$ is determined by the shape of the multi-field potential, and the amplitude of the isocurvature fluctuation during inflation.

Quantum tunneling during inflation: The radius of the bubble is of order $R_b = R_0\exp(-N_b)$, where $R_0$ is the radius of the current observable universe, and $N_b$ is the e-folding number from the beginning of observable inflation to the tunneling event. At late times during or after inflation, the tension of the bubble needs to vanish, in order that the bubble size is under control. The vanishing of the tension maybe realized by coupling the tunneling field to other dynamical fields, and let the late time evolution of the dynamical fields minimize the bubble tension. The number density of the bubble $n_b$ is determined by the tunneling rate. Or alternatively, if the energy difference between the false and true vacua get reduced due to dynamical mechanisms (while the bubble wall tension remains), the bubble wall can collapse and then disappear. 

Quantum tunneling after inflation: Similar to tunneling during inflation, at a later time, either the tension of the bubble needs to vanish, or the vacuum energy difference needs to vanish. Thus, the bubble expands for a period of time close to the speed of light, and the comoving size of the bubble is determined by the Hubble horizon size when the bubble wall disappears or collapses.

\section{Hawking radiation and intrinsic luminosity}
In this part, we briefly review the Hawking radiation and intrinsic luminosity of an individual PBH and refer to \cite{Carr:2009jm, Ukwatta:2015iba} for comprehensive studies.
It was found in \cite{Hawking:1974rv, Hawking:1974sw} that a black hole could emit particles similar to the black-body radiation, with energies in the range $(E,E + \mathrm{d}E)$ at a rate
\be \label{emission_rate}
\frac{\mathrm{d}^2 N}{\mathrm{d}t \mathrm{d}E} = \frac{1}{2 \pi} \frac{\Gamma_s(E,M)}{e^{8 \pi G M E} - (-1)^{2s}} ~,
\ee
per particle degree of freedom (e.g. spin, electric charge, flavor and color). Here $M$ is the mass of the black hole, $s$ is the particle spin. In contrast to the astrophysical black holes, PBHs collapsing from the overlarge primordial density perturbations could be small enough for Hawking radiation to be significant. The high-energy particles radiated from PBHs could influence various physical processes in the early Universe. Thus, one can impose evaporation constraints on PBH initial or current abundance via relevant observations, such as Big Bang Nucleosynthesis, CMB and gamma-ray observations. (The detailed discussions can see Refs. \cite{Carr:2009jm, Carr:2020gox} and the references therein). According to the radiation rate \eqref{emission_rate}, the black hole temperature can be defined as
\be \label{BH_Temperature}
T_\text{BH} = \frac{1}{8 \pi G M} \simeq 1.06 \times M_{10}^{-1} ~\text{TeV} ~,
\ee
where $M_{10}$ is related to the black hole mass $M \equiv M_{10} \times 10^{10} ~\text{g}$. And $\Gamma_s(E,M)$ is the dimensionless absorption coefficient which accounts for the probability that the particle would be absorbed if it were incident in this state on the black hole. The functional expressions of $\Gamma_s(E,M)$ for massless and massive particles can be found in Refs. \cite{Page:1976df, Page:1976ki, Page:1977um}. Hawking temperature \eqref{BH_Temperature} tells us that a smaller black hole is much hotter than a larger black hole, and the emission is also stronger. Note that we adopt the assumption that a black hole has no charge or angular momentum, which is reasonable since charge and angular momentum would also be lost through quantum emission on a shorter time scale than the mass loss time scale \cite{MacGibbon:1990zk, MacGibbon:1991tj}; extension to the charged and rotational black holes is straightforward \cite{Page:1976df, Page:1976ki, Page:1977um}. Since a black hole continuously emits particles, its mass decreases while the temperature goes up. The approximate formula for the mass loss rate can be written as \cite{MacGibbon:1990zk, Carr:2009jm}
\be \label{massloss_rate}
\frac{\mathrm{d} M_\text{10}}{\mathrm{d} t} \simeq - 5.34 \times 10^{-5} \phi(M) M_{10}^{-2} ~~ \text{s}^{-1} ~,
\ee
where $\phi(M)$ measures the number of emitted particle species and is normalized to unity for the black holes with $M \gg 10^{17}$ g, emitting only massless photons, three generations of neutrinos and graviton. The relativistic contributions to $\phi(M)$ per degree of particle freedom are $\phi_{s=0} = 0.267, \phi_{s=1} = 0.060, \phi_{s=3/2} = 0.020, \phi_{s=2} = 0.007, \phi_{s=1/2} = 0.147 ~(\text{neutral}), \phi_{s=1/2} = 0.142 ~(\text{charge} \pm e)$ \cite{MacGibbon:1990zk}.
Integrating the mass loss rate \eqref{massloss_rate} over time then gives the lifetime of a black hole
\be
\tau \sim 407 \l( \frac{\phi(M)}{15.35} \r)^{-1} M_{10}^3 ~~\text{s} ~.
\ee
If we sum up the contributions from all the particles in the Standard Model up to 1 TeV, corresponding to $M_{10} \sim 1$, this gives $\phi(M) = 15.35$. The mass of a PBH evaporating at $\tau$ after Big Bang is given by \cite{Carr:2009jm}
\be
M \simeq 1.35 \times 10^9 \l( \frac{\phi(M)}{15.35} \r)^{1/3} \l( \frac{\tau}{1 \text{s}} \r)^{1/3} ~\text{g} ~.
\ee
Thus, the mass of a PBH evaporating at present is roughly $M_* \simeq 5.1 \times 10^{14} ~\text{g}$ (corresponding to $T_\text{BH} = 21 ~\text{MeV}$).

Here, we adopt a standard emission picture that a black hole emits only those particles which appear elementary on the scale of the radiated energy (or equivalently the black hole size) \cite{MacGibbon:1990zk}. The emitted particles could form composite particles after emission. A black hole should emit all elementary particles whose rest masses are less than or of the order of $T_\text{BH}$. The spectra of the particles emitted through the lifetime of PBHs is calculated from the BlackHawk code \cite{Arbey:2019mbc}. When $T_\text{BH}$ increases, the black hole initially directly emits only photons (and gravitons), then neutrinos, electrons, muons and eventually direct pions join in the emission as $T_\text{BH}$ surpasses successive particle rest mass thresholds. Once the black hole temperature exceeds QCD energy scale $\Lambda_\text{QCD} = 250 - 300~\text{MeV}$, the particles radiated can be regarded as asymptotically free, leading to the emission of quarks and gluons. After their emission, quarks and gluons fragment into further quarks and gluons until they cluster into the observable hadrons including protons and antiprotons, electrons, and positrons. Since there are 12 quark degrees of freedom per flavor and 16 gluon degrees of freedom, one would expect the emission rate (i.e., the value of $\phi$) to increase suddenly once the QCD temperature is reached. Thus, Hawking radiation is dominated by the decay of QCD particles when the PBHs masses falls below $M_q \simeq 0.4 M_* \simeq 2 \times 10^{14}$ g \cite{MacGibbon:1990zk, MacGibbon:1991tj, Carr:2009jm}.

\begin{figure}[ht]
	\centering
	\includegraphics[width=\linewidth]{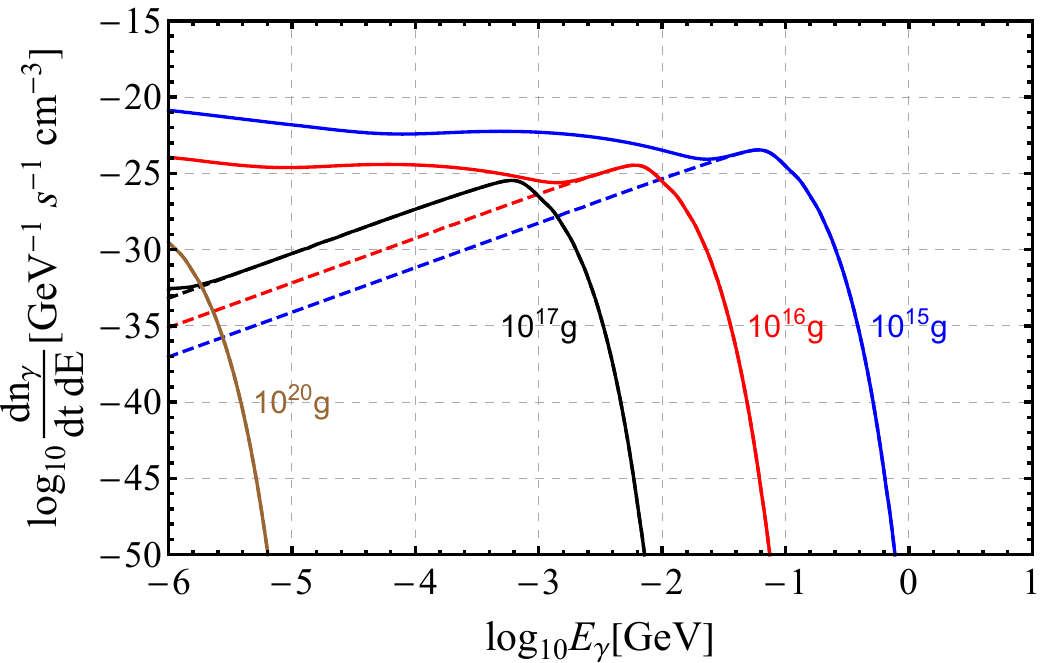}
	\caption{Instantaneous emission rate of photons per physical cm$^3$ from PBHs with various horizon masses: $10^{15}, 10^{16}, 10^{17}, 10^{20}$ g, and the corresponding current energy fraction is set to $\tilde{f}_{\text{PBH}} = 1$. The dashed and the solid ones represent the instantaneous primary and total (primary + secondary) emission rates, respectively. We used the open source code BlackHawk \cite{Arbey:2019mbc} to calculate the above photon radiated spectra.}
	\label{fig:hawkD}
\end{figure}

As discussed above, particles injected from a PBH have two components: the primary component, which is the direct Hawking emission; the secondary component, which comes from the decay of gauge bosons or heavy leptons and the hadrons produced by fragmentation of primary quarks and gluons \cite{Carr:2009jm}. For photons, we have
\be
\frac{\mathrm{d} \dot{N}}{\mathrm{d} E} (E, M)
= \frac{\mathrm{d} \dot{N}^\text{pri}}{\mathrm{d} E}(E, M)
+ \frac{\mathrm{d} \dot{N}^\text{sec}}{\mathrm{d} E}(E, M) ~,
\ee
with similar expressions to other particles. Fig.~\ref{fig:hawkD} plots the instantaneous emission rate of photons per physical cm$^3$ for PBHs with various horizon masses $10^{15}, 10^{16}, 10^{17}, 10^{20}$ g, and the corresponding energy fraction is set to $\tilde{f}_{\text{PBH}} = 1$. As we expect, the primary photons instantaneous spectrum dominates the high-tail of radiated spectra for the heavy PBHs (i.e., $M > 10^{15}$). This figure is similar to the Fig.~1 in Ref. \cite{Carr:2009jm} which shows the instantaneous emission rate of photons for four typical black hole masses.

The intrinsic luminosity of Hawking radiation from a PBH bubble is given by
\be \label{luminosity_HR}
L(E,t) = E { \mathrm{d}^2 n_\gamma \over \mathrm{d}t \mathrm{d}E} V \mathrm{d}E
\simeq
E^2 { \mathrm{d}^2 n_\gamma \over \mathrm{d}t \mathrm{d}E} V
\ee
with dimensions $\text{GeV}~\text{s}^{-1}$. Note that we used the approximation of the energy interval in the logarithmic scale $\mathrm{d}E \simeq E$. Here, $E$ is the emitted photon energy from the PBH bubble. The nearly time-independent behavior of $\mathrm{d}^2 n_\gamma / \mathrm{d}t \mathrm{d}E$ during early times, as shown in Fig. ~1, is due to the fact that the major contribution to the mass integral in Eq. (1) is made by the low-mass range above the low bound $M_\text{min}$. At the early time, the low-mass range barely changes, which leads to the time-independent instantaneous emission rates $\mathrm{d}^2 n_\gamma / \mathrm{d}t \mathrm{d}E$ in Fig.~1. As the PBHs evaporate, the low bound $M_\text{min}$ would go up, successively reaching the final evaporation stage of low-mass PBHs, the emission rates thus bump up at a later time. Finally, the emission rate would fade out on account of the evaporation of the low-mass range.

\section{Formation and merger rate of PBH binaries}
In this part, we briefly review the formation of PBH binaries and its merger rate. For more details, we would like to refer to \cite{Ali-Haimoud:2017rtz, Chen:2018czv}.

The PBH binary forms when two neighboring black holes are close enough and decouple from the Hubble flow. Given the equation of motion of two-point masses $M$ at rest with initial separation $x$, the proper separation $r$ along the axis of motion evolves as
\be\label{eom}
\ddot{r} - (\dot{H} + H^2) r + \frac{2M}{r^2}\frac{r}{|r|} = 0~,
\ee
where dots represent the differentiation with respect to proper time. In order to describe the early-Universe evolution, we use $s \equiv a/a_{\mathrm{eq}}$ the scale factor normalized to unity at matter-radiation equality to express the Hubble parameter as $H(s) = (8 \pi \rho_{\mathrm{eq}}/3)^{1/2} h(s)$, where $h$ is defined as $h(s) \equiv \sqrt{s^{-3} + s^{-4}}$ and $\rho_{\mathrm{eq}}$ is the matter density at equality. Then, we can rewrite Eq.~\eqref{eom} by introducing $\chi \equiv r/x$ as
\be
\chi'' + \frac{sh' + h}{s^2 h}(s \chi' - \chi) + \frac{1}{\lambda}\frac{1}{(sh)^2}\frac{1}{\chi^2}\frac{\chi}{|\chi|} = 0~,
\ee
where primes denote differentiation with respect to $s$. Here, the dimensionless parameter $\lambda$ is defined as $\lambda \equiv 4 \pi \rho_{\mathrm{eq}} x^3/3 M$. The initial condition is given in the condition that the two neighboring black holes follow the Hubble flow $\chi(s) = s$, the initial conditions are
\be
\chi(0) = 0~,~~~~~~~\chi'(0) = 1~.
\ee
Then, the numerical solution in \cite{Ali-Haimoud:2017rtz} shows that the binary effectively decouples from the Hubble flow at $s \approx \lambda/3$. The corresponding redshift is
\be
z = \frac{3(1 + z_{\mathrm{eq}})}{\lambda} - 1~.
\ee
In order to get the merger rate of PBH binaries with mass distribution $P(m)$ at cosmic time $t$, we follow \cite{Chen:2018czv} and define the binned mass distribution $P(m)$ and mass interval $\Delta$, which follows
\be
\sum_{m_{\mathrm{min}}}^{m_{\mathrm{max}}} P(m)\Delta = 1~.
\ee
The average distance between two nearby black holes is
\be
\langle x_{ij} \rangle = (\bar{x}_{i}^{-3} + \bar{x}_{j}^{-3})^{-1/3} = \mu_{ij}^{1/3}\bar{x}_{ij}~,
\ee
where, $\mu_{ij}$ and $\bar{x}_{ij}$ are defined as
\be
\mu_{ij} = \frac{2 m_i m_j f_b}{m_b f (P(m_j)m_i + P(m_i)m_j)}~,~~~\bar{x}_{ij}^3 = \frac{3}{8 \pi}\frac{m_b}{\rho_{\mathrm{eq}} f_b \Delta}~,
\ee
where, $f_b = f(P(m_i) + P(m_j))$ and $m_b = m_i + m_j$. Here, $f$ is the fraction of PBH energy density in matter. The relation of $f$ and $\tilde{f}_{\mathrm{PBH}}$ is $f \approx 0.85 \tilde{f}_{\mathrm{PBH}}$.
After the binary forms, GWs are emitted from the PBH binary. The coalescence time of the PBH bianry is given in \cite{Peters:1964zz} by
\be \label{coal_time}
t = \frac{3}{85}\frac{a^4}{G^3 m_1 m_2 M}j^7~.
\ee
From Eq.~\eqref{coal_time}, the merger rate of PBH binaries can be obtained from the initial semi-major axis $a$ distribution and the initial dimensionless angular momentum $j$ distribution. The initial semi-major axis $a$ after the binary formation is numerically given in \cite{Ali-Haimoud:2017rtz},
\be \label{major_axis}
a \approx 0.1 \lambda x = \frac{0.1 \bar{x}_{ij}}{f_b \Delta}X^{4/3},
\ee
where, $X \equiv x^3/\bar{x}_{ij}^3$.
Therefore, the initial semi-major axis distribution is determined by the separation $x$ distribution, we follow \cite{Ali-Haimoud:2017rtz, Chen:2018czv} that assuming PBHs follow a random distribution, the probability distribution of the separation $x$ is
\be
\frac{dP}{dX} = \mu_{ij}^{-1} e^{-X\frac{4\pi}{3}\bar{x}_{ij}^3n_{T}}~,
\ee
where $n_{T} \equiv \tilde{f}_{\mathrm{PBH}} \rho_{\mathrm{DM}}(1+z_{\mathrm{eq}})^3\int_0^{\infty} \frac{P(m)}{m}dm$. Considering fixed $X$, the dimensionless angular momentum $j$ can be given by Eq.~\eqref{coal_time} and Eq.~\eqref{major_axis},
\be
j(t;X) = \left( \frac{3}{85} \frac{G^3 m_1 m_2 M (f_b \Delta)^4}{(0.1 \bar{x}_{ij})^4 X^{16/3}} t \right)^{1/7}~.
\ee
The differential probability distribution of $(X, t)$ is given by
\be \label{xt_dis}
\begin{aligned}
\frac{d^2P}{dXdt} 
=& \frac{dP}{dX}\left(  \frac{\partial j}{\partial t}\frac{dP}{dj} \bigg|_X \right)_{j(t;X)}
\\=& \frac{\mu_{ij}^{-1}}{7t}e^{-X\frac{4\pi}{3}\bar{x}_{ij}^3n_{T}}\mathcal{P}(j/j_X)~,
\end{aligned}
\ee
where $j_X = 0.5 f X/ f_b \Delta$, $\mathcal{P}(j/j_X) = (j/j_X)^2/(1 + j^2/j_X^2)^{3/2}$. Integrating Eq.~\eqref{xt_dis} gives the merger time probability distribution
\be
\frac{dP}{dt} = \frac{\mu_{ij}^{-1}}{7t}\int dX e^{-X\frac{4\pi}{3}\bar{x}_{ij}^3n_{T}}\mathcal{P}(j/j_X)~.
\ee
Then, the comoving merger rate $R_{ij}$ for binary system at time $t$ is
\be
R_{ij}(t) =  \rho_{\mathrm{PBH}}\mathrm{min}\left(\frac{P(m_i)}{m_i},\frac{P(m_j)}{m_j}\right)\Delta \frac{dP}{dt}~.
\ee
The merger rate for the whole PBH distribution can be obtained by summarizing all the binaries system,
\be \label{EventRate}
R(t) = \!\!\!\!\! \sum_{\substack{0<m_i<m_j\\0<m_j<\infty}}\!\!\!\!\! \rho_{\mathrm{PBH}}\mathrm{min}\left(\frac{P(m_i)}{m_i},\frac{P(m_j)}{m_j}\right)\Delta \frac{dP}{dt}~.
\ee
%


\section{GW spectrum}
\begin{figure}[ht]
\centering
\includegraphics[width=8cm]{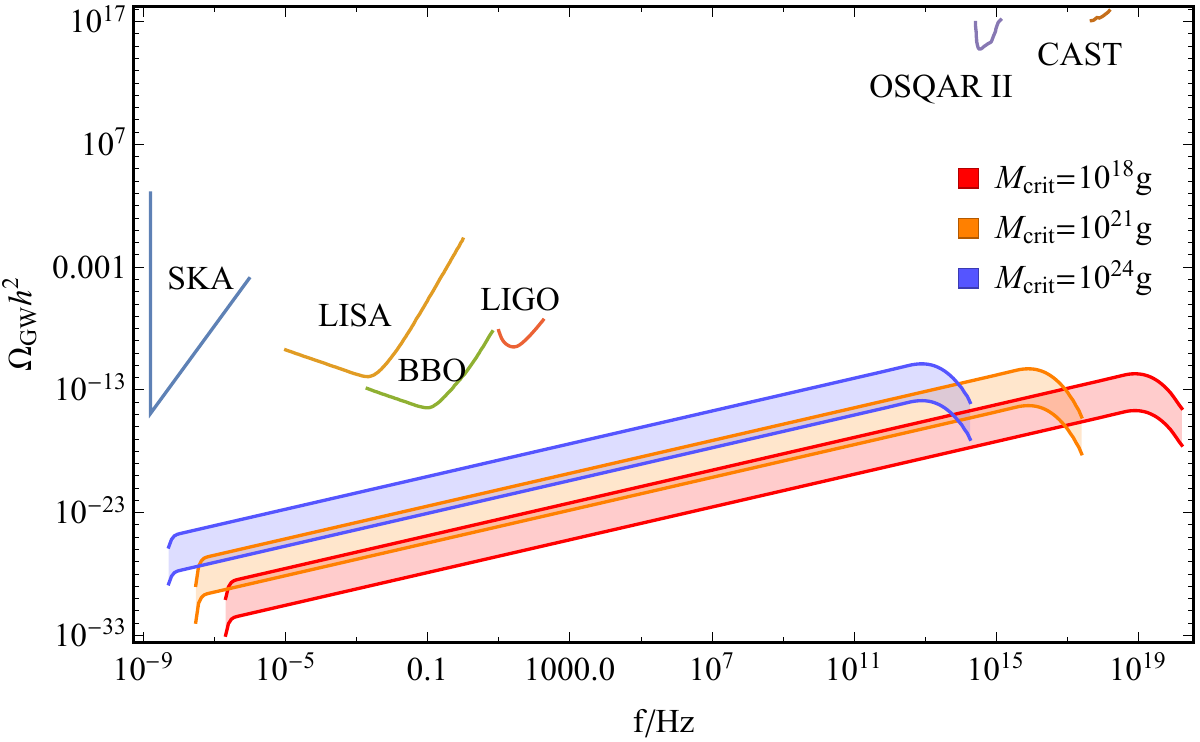}
\caption{\label{fig:HawkingGWspectrum}
	The GW spectrum of a single PBH stellar bubble. The $\tilde{f}_{\mathrm{PBH}} = 1$ in PBH stellar bubble and redshift of bubble is set as $z = 0.01$. The extended mass distribution of PBH is lognormal distribution with $\sigma = 1$ and $M_\text{pk} = 10^{18} \mathrm{g}, 10^{21} \mathrm{g}, 10^{24}\mathrm{g}$ for red, orange, blue shadow region respectively. The lower solid line and upper solid line denote the total PBH mass in bubble which is set as $10^{45} \mathrm{g}$ and the $10^{48} \mathrm{g}$ respectively. Sensitivity curve below $1000 \mathrm{Hz}$ from SKA, LISA, BBO and LIGO are plotted. Sensitivity curve in ultra-high frequency range from OSQAR \cite{Pugnat:2013dha} and CAST \cite{Anastassopoulos:2017ftl} are also plotted.}
\end{figure}

The GW energy flux from a distant source can be expressed as the following form \cite{Phinney:2001di},
\be
S(t) = \frac{L_{\mathrm{GW}}(t)}{4 \pi d_{L}^2}~,
\ee
where $L_{\mathrm{GW}}(t)$ is the GW luminosity measured in the PBH rest frame. Then, integrating $S(t)$ over the PBH binary evolution gives the observed redshifted GW energy, which is
%
%
\be
\int_{-\infty}^{\infty} S(t)dt = \frac{1 + z}{4 \pi d_L^2}\int_0^{\infty}\frac{dE_{\mathrm{GW}}}{df_r}df_r~.
\ee
Here the relation of time measured in the rest frame of source and observed frame is $t_r = t/(1+z)$. The observed energy density can be expressed as
\be
\begin{aligned}
\rho_{\mathrm{GW}} 
=& \int_0^{\infty} \Omega_{\mathrm{GW}}(f)\rho_c \frac{df}{f}
\\=&
\int_{z_{\mathrm{min}}}^{z_{\mathrm{max}}} \frac{1 + z}{4 \pi d_L^2} \left( \int_0^{\infty} f_r \frac{dE_{\mathrm{GW}}}{df_r} \frac{df}{f}\right) \frac{dN}{dt dz} dz~.
\end{aligned}
\ee
Here, $dN/dtdz$ is the number of merger events which occur in $dt$ between redshift $z$ and $z + dz$. Therefore, $\Omega_{\mathrm{GW}}$ can be written as
\be \label{OmegaGW1}
\Omega_{\mathrm{GW}}(f) = \frac{1}{\rho_c}\frac{1}{4 \pi d_L^2} f_r \frac{dE_{\mathrm{GW}}}{df_r} \int_{z_{\mathrm{min}}}^{z_{\mathrm{max}}} (1 + z) \frac{dN}{dt dz}dz~.
\ee
Compare with the cosmic distance, the comoving size of PBH stellar bubble is relatively small, so that $\Delta z = z_{\mathrm{max}} - z_{\mathrm{min}}$ is tiny. Eq.~\eqref{OmegaGW1} can be written as
\be \label{OmegaGW2}
\Omega_{\mathrm{GW}}(f) = \frac{1}{\rho_c}\frac{1}{4 \pi d_L^2}f_r \frac{dE_{\mathrm{GW}}}{df_r}\frac{dN}{dt_r}
= \frac{1}{\rho_c}\frac{1}{4 \pi d_L^2}f_r \frac{dE_{\mathrm{GW}}}{df_r}R~.
\ee
From the Eq.~\eqref{OmegaGW2}, we can get the GW spectrum from the single PBH stellar bubble in Fig.~\ref{fig:HawkingGWspectrum}
.

In Fig.~\ref{fig:HawkingGWspectrum}, the GW energy density from the PBH stellar bubble with small peak mass is weak in detection. However, the peak energy density of PBH stellar bubble weakly depends on the peak mass. In Eq.~(5), the peak energy radiation occurs at $f_r = f_2$, so
\be
f_r \frac{dE_{\mathrm{GW}}}{df_r} = M \eta \frac{(a_2 \eta^2 + b_2 \eta +c_2)^{5/3}}{a_1 \eta^2 + b_1 \eta +c_1}.
\ee
In Eq.~\eqref{EventRate}, $R \sim M^{-32/37}$, see \cite{Sasaki:2016jop,Chen:2018czv,Ding:2019tjk} for details. As the result, the peak radiation power in the rest frame of PBH stellar bubble is
\ba \label{mcdependence}
R f_r \frac{dE_{\mathrm{GW}}}{df_r} &\sim & M^{5/37} \eta \frac{(a_2 \eta^2 + b_2 \eta +c_2)^{5/3}}{a_1 \eta^2 + b_1 \eta +c_1}\\\nonumber
&\sim & M_\text{pk}^{5/37} (\gamma_1 + \gamma_2)^{5/37} \eta \frac{(a_2 \eta^2 + b_2 \eta +c_2)^{5/3}}{a_1 \eta^2 + b_1 \eta +c_1}~.
\ea
Here, we define mass ratio $\gamma_i \equiv m_i/M_\text{pk}$. The $\gamma$ and $\eta$ can be shown that they are independent on the $M_\text{pk}$ in lognormal distribution as following,
\bl
\psi_\text{LN}(M) dM &=& {\tilde{f}_{\text{PBH}} \over \sqrt{2 \pi} \sigma M} \exp\Big[  - { \ln^2(M/M_\text{pk}) \over 2 \sigma^2 }  \Big]dM \\ \nonumber
&=& { \tilde{f}_{\text{PBH}} \over \sqrt{2 \pi} \sigma \gamma M_\text{pk}} \exp\Big[  - { \ln^2(\gamma) \over 2 \sigma^2 }  \Big] M_\text{pk} d\gamma~.
\el
Therefore, the $\gamma$ distribution is
\be
\psi_\text{LN}(\gamma) d\gamma ={\tilde{f}_{\text{PBH}} \over \sqrt{2 \pi} \sigma \gamma} \exp\Big[  - { \ln^2(\gamma) \over 2 \sigma^2 }  \Big] d\gamma~,
\ee
which is independent with $M_\text{pk}$. $\eta$ can be expressed as
\be
\eta = \frac{m_1 m_2}{(m1 + m2)^2} = \frac{\gamma_1 \gamma_2}{(\gamma_1 + \gamma_2)^2}~,
\ee
which is also independent with $M_\text{pk}$. Therefore Eq.~\eqref{mcdependence} shows that the peak energy density depends on the $M_\text{pk}^{5/37}$, which weakly depends on the peak mass in PBH distribution.

\bibliographystyle{apsrev4-1}
\bibliography{pbh_bubble}
\end{document}